\documentclass[nonacm,sigplan]{acmart}

\usepackage[]{hyperref}
\usepackage[normalem]{ulem}
\usepackage{graphicx}
\usepackage{algorithm}
\usepackage{algpseudocode}
\usepackage{tikz}
\usetikzlibrary{arrows}
\usepackage{enumerate}
\usepackage{enumitem}
\usepackage{comment}
\usepackage{pifont}
\usepackage{hyperref}

\usepackage{booktabs}
\usepackage{tabularx}
\usepackage{makecell}

\usepackage{xcolor}

\usepackage{caption}
\usepackage{subcaption}
\usepackage[normalem]{ulem}

\newcommand{\dcircle}[1]{\ding{\numexpr181 + #1}}

\newcommand{\cusz}{\textsc{cuSZ}}
\newcommand{\cuzfp}{cuZFP}

\newcommand{\cudamgard}{MGARD-GPU}
\renewcommand{\cusz}{\textsc{cuSZ}}  

\definecolor{bygreen}{RGB}{57, 151, 92}
\newcommand{\cmark}{\textcolor{bygreen}{\ding{51}}}%
\newcommand{\xmark}{\textcolor{red}{\ding{55}}}%

\newcommand{\thiswork}{\textsc{BMQSim}}

\usepackage[many]{tcolorbox}    	
\usepackage{multicol}               

\definecolor{main}{HTML}{283618}    
\definecolor{sub}{HTML}{FBF8CC}     

\definecolor{mainI}{HTML}{283618}    
\definecolor{subI}{HTML}{FFCFD2}     

\newtcolorbox{boxK}{
    sharpish corners, 
    boxrule = 0pt,
    toprule = 4.5pt, 
    enhanced,
    fuzzy shadow = {0pt}{-2pt}{-0.5pt}{0.5pt}{black!35} 
}

\newtcolorbox{boxG}{
    enhanced,
    boxrule = 0pt,
    colback = sub,
    borderline west = {1pt}{0pt}{main}, 
    borderline west = {0.75pt}{2pt}{main}, 
    borderline east = {1pt}{0pt}{main}, 
    borderline east = {0.75pt}{2pt}{main}
}

\newtcolorbox{boxI}{
    enhanced,
    boxrule = 0pt,
    colback = subI,
    borderline west = {1pt}{0pt}{mainI}, 
    borderline west = {0.75pt}{2pt}{mainI}, 
    borderline east = {1pt}{0pt}{mainI}, 
    borderline east = {0.75pt}{2pt}{mainI}
}

\newcommand{\SEC}{\S}

\begin{document}

\title{Overcoming Memory Constraints in Quantum Circuit Simulation with a High-Fidelity Compression Framework}

\author{Boyuan Zhang}
\affiliation{%
  \institution{Indiana University}
  \city{Bloomington}
  \state{Indiana}
  \country{USA}}
\email{bozhan@iu.edu}

\author{Bo Fang*}
\affiliation{%
  \institution{Pacific Northwest National Laboratory}
  \city{Richland}
  \state{Washington}
  \country{USA}}
\email{bo.fang@pnnl.gov}

\author{Fanjiang Ye}
\affiliation{%
  \institution{Indiana University}
  \city{Bloomington}
  \state{Indiana}
  \country{USA}}
\email{fanjye@iu.edu}

\author{Yida Gu}
\affiliation{%
  \institution{University of Chinese Academy of Sciences}
  \city{Beijing}
  \country{China}}
\email{guyida21@mails.ucas.ac.cn}

\author{Nathan Tallent}
\affiliation{%
  \institution{Pacific Northwest National Laboratory}
  \city{Richland}
  \state{Washington}
  \country{USA}}
\email{Nathan.Tallent@pnnl.gov}

\author{Guangming Tan}
\affiliation{%
  \institution{Institute of Computing Technology, Chinese Academy of Sciences}
  \city{Beijing}
  \country{China}}
\email{tgm@ict.ac.cn}

\author{Dingwen Tao*}
\affiliation{%
  \institution{Institute of Computing Technology, Chinese Academy of Sciences}
  \city{Beijing}
  \country{China}}
\email{taodingwen@ict.ac.cn}
\thanks{* Co-corresponding author}
\begin{abstract}
Full-state quantum circuit simulation requires exponentially increased memory size to store the state vector as the number of qubits scales, presenting significant limitations in classical computing systems. Our paper introduces \textit{\thiswork}, a novel state vector quantum simulation framework that employs lossy compression to address the memory constraints on graphics processing unit (GPU) machines. \thiswork{} effectively tackles four major challenges for state-vector simulation with compression: frequent compression/decompression, high memory movement overhead, lack of dedicated error control, and unpredictable memory space requirements. Our work proposes an innovative strategy of circuit partitioning to significantly reduce the frequency of compression occurrences. We introduce a pipeline that seamlessly integrates compression with data movement while concealing its overhead. Additionally, \thiswork{} incorporates the first GPU-based lossy compression technique with point-wise error control. Furthermore, \thiswork{} features a two-level memory management system, ensuring efficient and stable execution. Our evaluations demonstrate that \thiswork{} can simulate the same circuit with over 10 times less memory usage on average, achieving fidelity over 0.99 and maintaining comparable simulation time to other state-of-the-art simulators.

\end{abstract}

\maketitle 
\pagestyle{plain} 

\section{Introduction}
\label{sec:introduction}
In recent years, quantum computing has proven effective in addressing key problems across various fields, such as machine learning \cite{lloyd2013quantum, biamonte2017quantum, schuld2019quantum}, quantum chemistry \cite{aspuru2005simulated}, optimization problems \cite{farhi2014quantum}, and financial modeling \cite{rebentrost2018quantum}. The advancement of quantum hardware aligns with the increasing impact of quantum computing. For instance, the state-of-the-art (SOTA) IBM Condor quantum system now supports 1,121 qubits, more than double of the 433 qubits supported by last year's Osprey quantum system \cite{gambetta2020ibm}. Executing quantum algorithms on real quantum computers, however, faces fundamental challenges. First, in the current Noisy Intermediate-Scale Quantum (NISQ) era \cite{preskill2018quantum}, noise interference in the hardware results in inaccurate measurement distribution. Second, designing new quantum algorithms requires iterative trials to verify, which is impractical on quantum computer platforms. Third, publicly available quantum computers (specifically those with a large number of qubits, e.g., > 16) are much less resourceful and usually reside in cloud services; hence, access to those machines is limited. Thus, quantum circuit simulation has become an essential approach for realizing the full potential of quantum computing \cite{jones2019quest}.
Running a full-state quantum circuit simulation (i.e., state-vector simulation) presents a formidable challenge: as the number of simulated qubits increases, the memory requirement grows exponentially. Several significant issues are associated with this:
(1) Simulating large quantum systems requires extensive memory capacity in classical systems. For instance, simulating a 48-qubit circuit would fully occupy the entire memory of Frontier (4.6 petabytes of DDR4 memory), the most advanced high-performance computing (HPC) machine currently available \cite{atchley2023frontier}.
(2) Even when the memory capacity requirement is met, accessing such HPC systems requires dedicated allocation, which is usually quite competitive due to high demand. Consequently, researchers in quantum computing are often constrained to work with much smaller machines, such as personal computers or local workstations that typically have only dozens of gigabytes of memory. This reliance severely restricts the ability to simulate large quantum systems, hindering scientific discovery.

While recent developments in state-vector simulators have made significant strides in performance improvement \cite{li2021sv, zhang2022uniq, zhang2021hyquas, hisvsim}, optimizing memory usage remains a largely overlooked area. Tensor network simulation is expected to address this issue \cite{pang2020efficient, markov2008simulating} by representing the quantum circuit using tensor structures and employing tensor contraction to compute the final state vector amplitudes. However, tensor network simulators face significant limitations when simulating highly entangled quantum circuits \cite{white1992density, ostlund1995thermodynamic}. For entanglement-heavier circuits, both the computational and memory overhead of tensor network simulators grow substantially. This restricts their applicability primarily to circuits that are shallow and exhibit low entanglement between qubits. For instance, using tensor network simulators to execute the Quantum Approximate Optimization Algorithm (QAOA) \cite{farhi2014quantum} and the Variational Quantum Eigensolver (VQE) \cite{peruzzo2014variational}, the most representative quantum algorithms in the NISQ era, faces significant limitations. In QAOA, tensor networks can only efficiently manage a limited number of layers \cite{lykov2021performance}, while an arbitrary number of operational layers is essential to increase effectiveness \cite{farhi2020quantum, guerreschi2019qaoa}. For VQE, the enormous number of gates and the level of qubit entanglement \cite{taube2006new} create impractical scenarios for tensor networks to solve. 

That said, state vector-based quantum circuit simulation offers generality and universal benefits for simulating complex quantum algorithms. To this end, relaxing the memory constraint for state vector simulation is the top priority task. In the classical HPC domain, data compression has proven effective in multiple scientific areas for memory reduction. Broadly speaking, compression techniques can be classified into lossy and lossless, based on the trade-offs between the error and compression ratio they introduce to the data. Compared to lossless compression, lossy compression tends to provide better compression rates \cite{zhang2023fz, zhang2023gpulz}, making it more suitable for high-memory burden scenarios like quantum simulation. Recent studies \cite{zhang2023fz, cusz2020, chen2021accelerating, huang2023cuszp} have developed error-bounded lossy compressors on GPUs, achieving a balance between compression ratios, high-quality data reconstruction, and performance. Incorporating these advanced compression algorithms into quantum simulation holds considerable promise for significantly reducing memory demands, thereby addressing the fundamental challenge in the field.

However, the direct application of a compression technique on state-vector simulations is inefficient and may result in low simulation fidelity. A prior study \cite{wu2019full} introduces a workflow that addresses this integration. The workflow starts with compressing the entire state vector. For each gate in the circuit, it breaks the compressed elements into blocks, decompresses each block, updates the state elements in the block, and then re-compresses it until all blocks are processed.
This design introduces several potential complications, particularly concerning the performance of the simulation and the fidelity of the quantum state. These issues encompass five primary domains:

\textbf{Challenge\dcircle{1}: Frequent Compression.}
Since the entire state vector needs to be updated when simulating each quantum gate, a large quantum system would require frequent compression and decompression operations on the critical path of the state vector simulation, introducing significant performance overhead. Moreover, lossy compression inherently introduces errors into the reconstructed data. When simulating deep quantum circuits, these errors accumulate and degrade the fidelity of the final results.

\textbf{Challenge\dcircle{2}: Memory Movement Overhead.}
To maximize the number of qubits supported by simulation and improve the simulation performance, the involvement of large memory space such as CPU memory and high-parallelism computing resources like GPUs is necessary. However, the data movement between the CPU and GPU to take advantage of computation acceleration incurs significant overhead.

\textbf{Challenge\dcircle{3}: Lack of Dedicated Error Control Scheme.}
Effective error control in lossy compression is essential, particularly for the point-wise relative error control scheme for state-vector simulation \cite{wu2019full}. The GPU-based compression processes would outperform their CPU-based alternatives and eliminate potential additional memory transfers between the CPU and GPU. However, current GPU-based lossy compressors do not incorporate such a scheme.

\textbf{Challenge\dcircle{4}: Unpredictable Memory Consumption of Compressed State Vectors.}
When handling large input data, lossy compressors often divide the data into smaller chunks for independent compression. However, the memory footprints of the compressed state vector chunks depend on the properties of the state vector, complicating the accurate assessment of whether the available memory will suffice for the simulation.

In response to these challenges, we introduce a novel state vector quantum simulation framework, \thiswork
, by efficiently integrating lossy compression techniques. This framework can break the memory limit to support the robust simulation of more qubits on GPU machines while maintaining high fidelity in simulation results by significantly reducing the frequency of compression with a novel circuit partition scheme. \thiswork{} is adaptable, allowing for easy integration into various simulators, enhancing its utility across different simulation backends.

Our paper makes the following contributions:
\begin{itemize}[leftmargin=1.3em]
\item We introduce a novel circuit partitioning strategy, effectively addressing low-fidelity and low-performance concerns of the compression-integrated simulation. This method divides the simulation process into discrete subtasks, each involving a partition of the circuit and corresponding elements of the state vector. This approach significantly reduces the frequency of compression and decompression operations, thereby maintaining exceptionally high simulation fidelity and significantly improving simulation time.

\item We propose an innovative workflow pipeline that concurrently executes (de)compression operations and data movement. This approach minimizes the perceived overhead in the simulation process by effectively \textit{hiding} these operations within the data transfer time frames.

\item We develop the first GPU-based point-wise error control mechanism in a lossy compressor. It offers adaptability to other compressors requiring absolute error control, marking a significant advancement in GPU-accelerated data compression.

\item We propose a two-level memory management system to address the challenge of unpredictable compressed state vector block sizes. It dynamically manages memory (de)allocation and uses the GPUDirect Storage technique to create an effective secondary memory buffer in an SSD, ensuring efficient memory utilization and enhanced operational stability.

\item Evaluations on various circuits demonstrate that \thiswork{} significantly enhances the capabilities of SOTA state-vector simulators by enabling the simulation of up to 14 additional qubits (on average 10 additional qubits) under the same memory constraints, while maintaining comparable simulation times to SOTA simulators.
\end{itemize}

This paper is organized as follows: \SEC\ref{sec:related} provides background information. \SEC\ref{sec:problem} analyzes the problem and discusses the issues of basic solutions. \SEC\ref{sec:design} details our design. Evaluation results are presented in \SEC\ref{sec:eval}. Finally, \SEC\ref{sec:conclusion} summarizes our findings and discusses future research directions.
\section{Background}
\label{sec:related}
In this section, we introduce state-vector simulation, floating-point data compression, and CUDA architecture.

\subsection{Principles of State-Vector Simulation}
\label{subsec:sv_update}
In quantum computing, a qubit, like a bit in classical computing, is the fundamental unit for computing. Unlike bits in traditional computing, a qubit can have many more states besides 0 and 1. A qubit \( |\psi\rangle \) is a two-level state that can be expressed as:
\[ |\psi\rangle = a_0|0\rangle + a_1|1\rangle \]
Here, \( a_0 \) and \( a_1 \) represent two complex amplitudes, where \( |a_0|^2 + |a_1|^2 = 1 \). 
The quantum state with \( n \) qubits can be described as a state vector containing \( 2^n \) complex amplitudes:
\[ |\psi\rangle = a_{0\cdots00}|0\cdots00\rangle + a_{0\cdots01}|0\cdots01\rangle + \cdots + a_{1\cdots11}|1\cdots11\rangle \]
This state also adheres to the condition 
\( \sum_i |a_i|^2 = 1 \). The subscripts of $a$ are the indices in binary format.
 In the computation of simulation, the state vector is often denoted as a column vector: 
\[
\begin{bmatrix} 
a_{0\cdots00} \\
a_{0\cdots01} \\
\vdots \\
a_{1\cdots11} 
\end{bmatrix}
\]

A quantum gate represents a unitary operation applied to qubit(s), and a series of quantum gates operating on a set of qubits forms a quantum circuit. Applying a gate to a qubit is equivalent to conducting a matrix multiplication of the gate unitary matrix and the elements in the state vector. These matrices modify the elements of the state vector corresponding to the target qubit(s). The most common types of gates are single-qubit gates and double-qubit gates. For a single-qubit gate (a \( 2 \times 2 \) matrix) applied to qubit \( k \), the operation is to multiply the matrix with two elements whose indices differ only in the \( k \) bit:
\begin{gather*}
\begin{bmatrix} a'_{e_{2^n-1}^1 \cdots 0^1_k \cdots e^1_0} \\ a'_{e_{2^n-1}^2 \cdots 1^2_k \cdots e^2_0} \end{bmatrix} = 
\begin{bmatrix} u_{11} & u_{12} \\ u_{21} & u_{22} \end{bmatrix} 
\begin{bmatrix} a_{e_{2^n-1}^1 \cdots 0^1_k \cdots e^1_0} \\ a_{e_{2^n-1}^2 \cdots 1^2_k \cdots e^2_0} \end{bmatrix},
\\
\forall \begin{bmatrix} a_{e^1} \\ a_{e^2} \end{bmatrix}, e^1_i = e^2_i \quad \text{for } 0 \leq i < 2^n \text{ and } i \neq k
\end{gather*}
where \( \begin{bmatrix} a_* \end{bmatrix} \) are the state vector amplitudes and \( \begin{bmatrix} u_* \end{bmatrix} \) is the unitary matrix of the applied gate.
Similarly, for a double-qubit gate (a \( 4 \times 4 \) matrix) applied to qubits \( q \) and \( k \), the matrix operation is:
\begin{gather*}
\begin{bmatrix} a'_{e^1_{2^n-1} \cdots 0^1_q \cdots 0^1_k \cdots e^1_0} \\ a'_{e^2_{2^n-1} \cdots 0^2_q \cdots 1^2_k \cdots e^2_0} \\ a'_{e^3_{2^n-1} \cdots 1^3_q \cdots 0^3_k \cdots e^3_0} \\ a'_{e^4_{2^n-1} \cdots 1^4_q \cdots 1^4_k \cdots e^4_0} \end{bmatrix} = 
\begin{bmatrix} u_{11} & u_{12} & u_{13} & u_{14} \\ u_{21} & u_{22} & u_{23} & u_{24} \\ u_{31} & u_{32} & u_{33} & u_{34} \\ u_{41} & u_{42} & u_{43} & u_{44} \end{bmatrix} 
\begin{bmatrix} a_{e^1_{2^n-1} \cdots 0^1_q \cdots 0^1_k \cdots e^1_0} \\ a_{e^2_{2^n-1} \cdots 0^2_q \cdots 1^2_k \cdots e^2_0} \\ a_{e^3_{2^n-1} \cdots 1^3_q \cdots 0^3_k \cdots e^3_0} \\ a_{e^4_{2^n-1} \cdots 1^4_q \cdots 1^4_k \cdots e^4_0} \end{bmatrix},
\\
\forall \begin{bmatrix} a_* \end{bmatrix},
e^1_i = e^2_i = e^3_i = e^4_i \quad \text{for } 0 \leq i < 2^n \text{ and } i \neq k, i \neq q
\end{gather*}
An important requirement for both single-qubit gates and double-qubit gates is that simulating a gate operation requires iterating through the entire state vector.


\subsection{Floating-Point Lossy Compression} 
\label{subsec:compression}

In the field of data compression, there are two main types: lossless and lossy compression. Lossless compression retains the original data perfectly, while lossy compression, in exchange for a higher compression ratio, incurs some loss of accuracy. The latter is suitable for scenarios where a certain level of data degradation is acceptable.

Recently, there have been significant advancements in lossy compression algorithms, particularly for floating-point scientific data. Prominent examples are SZ \cite{sz16, sz17, liang2018error, sz3}, ZFP \cite{zfp}, MGARD \cite{mgard, liang2021mgard+}, and TTHRESH \cite{ballester2019tthresh}. These algorithms are distinct from traditional lossy compressors for images/videos, as they feature precise error-controlling schemes. These schemes allow for control over the level of accuracy in reconstructed data and further data analysis.

With the rise of GPU-based systems, compatible versions of these compressors, such as \cusz{} \cite{cuSZ, cusz2020}, \cuzfp{} \cite{cuZFP}, and \cudamgard{} \cite{chen2021accelerating}, have been developed using CUDA \cite{sanders2010cuda}. Furthermore, new GPU-oriented lossy compressors like FZ-GPU \cite{zhang2023fz}, bitcomp \cite{nvcomp}, and cuSZp \cite{huang2023cuszp} have emerged. These GPU versions typically offer higher compression throughputs than their CPU counterparts.

However, a gap remains in current GPU compressors: most only support absolute error control or fixed-rate modes. The former keeps the maximum error within a user-defined limit, while the latter targets a specific compression ratio. A critical missing feature is a point-wise relative error control scheme, vital for state-vector simulation to ensure high fidelity \cite{wu2019full}.









\subsection{CUDA Memory Architecture}
\label{subsec:gpu}
The increasing adoption of GPUs as the main accelerators of high-performance computing tasks is primarily due to their superior parallel computation capabilities. Within the CUDA architecture \cite{sanders2010cuda}, a widely used programming model for GPUs, processing units are organized into threads. These threads are grouped into blocks, which are then organized into a grid structure. GPUs typically feature on-chip memory, or device memory, which is usually much less abundant compared to CPU memory or main memory. 



Most applications initialize memory allocation on the CPU and then copy the data to the GPU for computation through PCIe. Therefore, asynchronous memory copy operations are crucial for enhancing data transfer efficiency between the CPU and GPU. Such operations enable GPU kernels (GPU processes) to run concurrently with memory copy tasks, optimizing data transfer efficiency. Recently, data copying can occur directly between SSDs and GPU memory. For the movement of data between SSDs and GPUs, the GPUDirect Storage (GDS) technique is vital. This technique allows GPUs to directly access data stored on SSDs, bypassing the CPU and thus enhancing performance.
\section{Feasibility Analysis}
\label{sec:problem}
In this section, we provide a detailed analysis of the solution developed in SC19-Sim \cite{wu2019full} to integrate data compression with state vector simulation and identify its shortcomings. \textit{For simplicity, from now on, we use single-qubit gates and binary representation of indices in all the following examples.}

The prior work \cite{wu2019full} proposed a basic solution of applying compression techniques in state vector simulation. This solution consists of two key designs: state vector partitioning and state vector updating.

\begin{figure}[t]
    \centering
    \includegraphics[width=\linewidth]{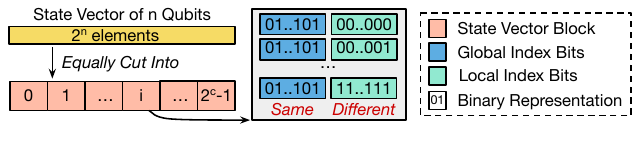}
    \caption{A demonstration of state vector partitioning. We refer to the higher \( c \) bits as the global index and the lower \( b \) bits as the local index.}
    \label{fig:sv_partition}
\end{figure}

\textbf{State Vector Partitioning.} To maximize flexibility and enable parallel execution of compression and simulation, SC19-Sim divides the state vector into blocks, which we term \uline{SV blocks}. A demonstration of the state vector partition is illustrated in Figure \ref{fig:sv_partition}. Assume that the state vector is divided into \uline{$2^c$} SV blocks, and each SV block contains \uline{$2^b$} state vector elements (i.e., amplitudes). Given an \uline{$n$}-qubit system, where $n = b + c$, the higher $c$ bits in the qubit index space are referred to as the \uline{global index}, while the lower $b$ bits are referred to as the \uline{local index}. A clear observation is that within each SV block, the global index remains the same, but the local index varies. Different SV blocks have different global indices.

\begin{figure}[t]
    \centering
    \includegraphics[width=\linewidth]{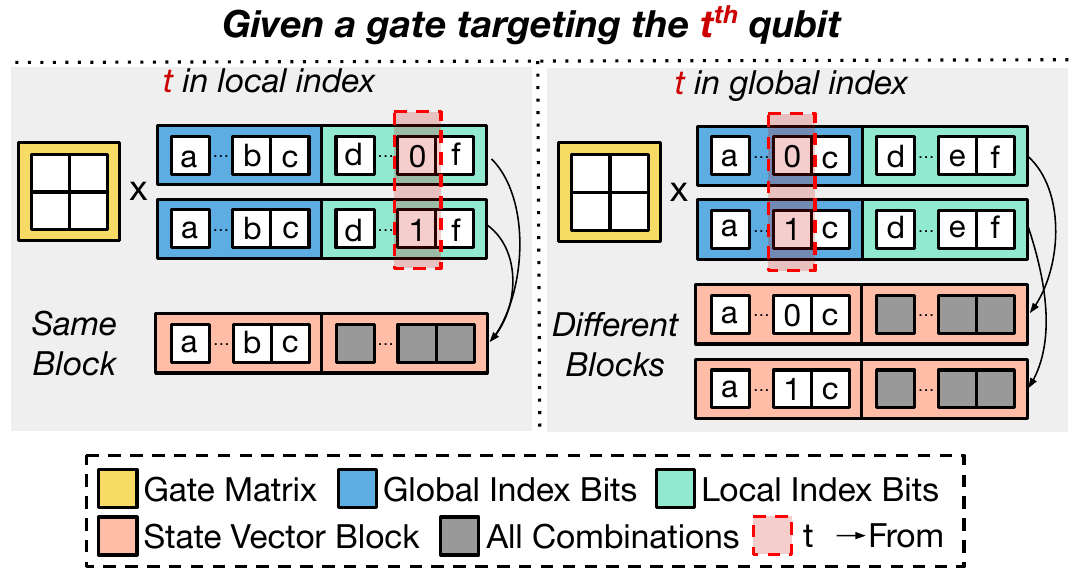}
    \caption{A demonstration of how the target qubit location influences amplitude updates. The same alphabet denotes the same value (0 or 1).}
    \label{fig:index_location}
\end{figure}

\textbf{State Vector Updates.} At the beginning of the simulation, the state vector (SV) blocks are compressed and stored in the system memory. During the simulation, each gate updates the entire state vector once (as discussed in \SEC\ref{subsec:sv_update}). This process involves decompressing every SV block, updating the amplitudes within it, and then recompressing it back into the system memory. Depending on whether the target qubit of the quantum gate is located in the global index or the local index, the updating process may involve either two separate SV blocks or a single SV block, as illustrated in Figure \ref{fig:index_location}. We summarize the updating rules as an observation.
\begin{boxG}
\textbf{Observation}: If the target qubit \( t_i \) is in the local index set, the amplitudes needed for matrix-vector multiplication are within the same block. Otherwise, the amplitudes are in different SV blocks, where their exact positions depend on the target qubit.
\end{boxG}

\textbf{Issues of the Basic Solution.} Based on this observation, the order of processing SV blocks in the simulation process may vary due to the order of the different target qubits of the gates in the circuit. Therefore, without careful design, SC19-Sim applies each gate sequentially to the state vector, requiring decompression and compression before and after updating the state vector amplitudes.

This design exposes several issues:
\dcircle{1} Since (de)compression is executed on a per-gate basis, fully decompressing all SV blocks for every gate operation significantly lowers performance. Moreover, as the circuit length (number of gates) increases, the number of lossy compression operations escalates, leading to an accumulation of errors and degradation of state fidelity.
\dcircle{2} The GPU is not leveraged, as the entire state vector is processed only by the CPU. Leveraging the parallel computing capability of GPUs can significantly improve performance. However, the intensive data transfer between CPU and GPU will heavily impact simulation efficiency.
\dcircle{3} The compression-introduced error is not controlled. Random errors introduced by compression will result in unguaranteed fidelity. Therefore, we need a specialized error control scheme to bound the fidelity.
\dcircle{4} The compression ratio is unpredictable during the simulation process. The simulation may halt midway due to insufficient memory space, necessitating a backup memory management system to prevent such interruptions.
\section{Design of \thiswork}
\label{sec:design}
\begin{figure*}[ht]
    \centering
    \includegraphics[width=\linewidth]{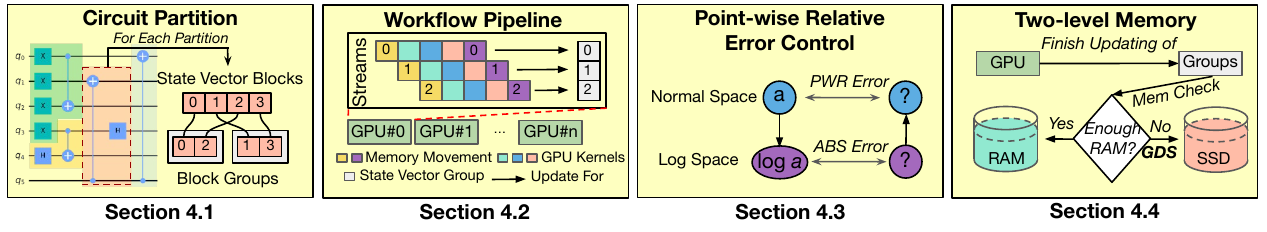}
    \caption{An overview of our proposed \thiswork{}.}
    \label{fig:highlevel_overview}
\end{figure*}

\textbf{Overview of \thiswork{}'s Design.} \thiswork{} is designed to simulate full-state quantum circuits with a smaller memory footprint to support more qubit systems. Figure \ref{fig:highlevel_overview} summarizes the key techniques implemented in \thiswork{} and the respective sections where they are discussed. Specifically, we introduce a specialized circuit partition approach (\SEC\ref{subsec:hardware-sv-partition}) to minimize the (de)compression frequency hence significantly improve the performance and increase the fidelity, addressing \dcircle{1}. We include a workflow design (\SEC\ref{subsec:pipeline-design}) to overlap compression/decompression, data movement between CPUs and GPUs, and computation, addressing \dcircle{2}. An error-controlled GPU compressor (\SEC\ref{subsec:gpu-compression}) is proposed to mitigate \dcircle{3}. Finally, we present a two-level memory management system (\SEC\ref{subsec:fall-back}) to solve issue \dcircle{4}.

\subsection{Optimal-Compression Circuit Partition}
\label{subsec:hardware-sv-partition}

As analyzed in Section~\ref{sec:problem}, the basic solution leads to frequent (de)compression because gates in the circuit require different access patterns on SV blocks due to different target qubits. This issue significantly impacts the simulation performance.

\begin{figure}[t]
    \centering
    \includegraphics[width=\linewidth]{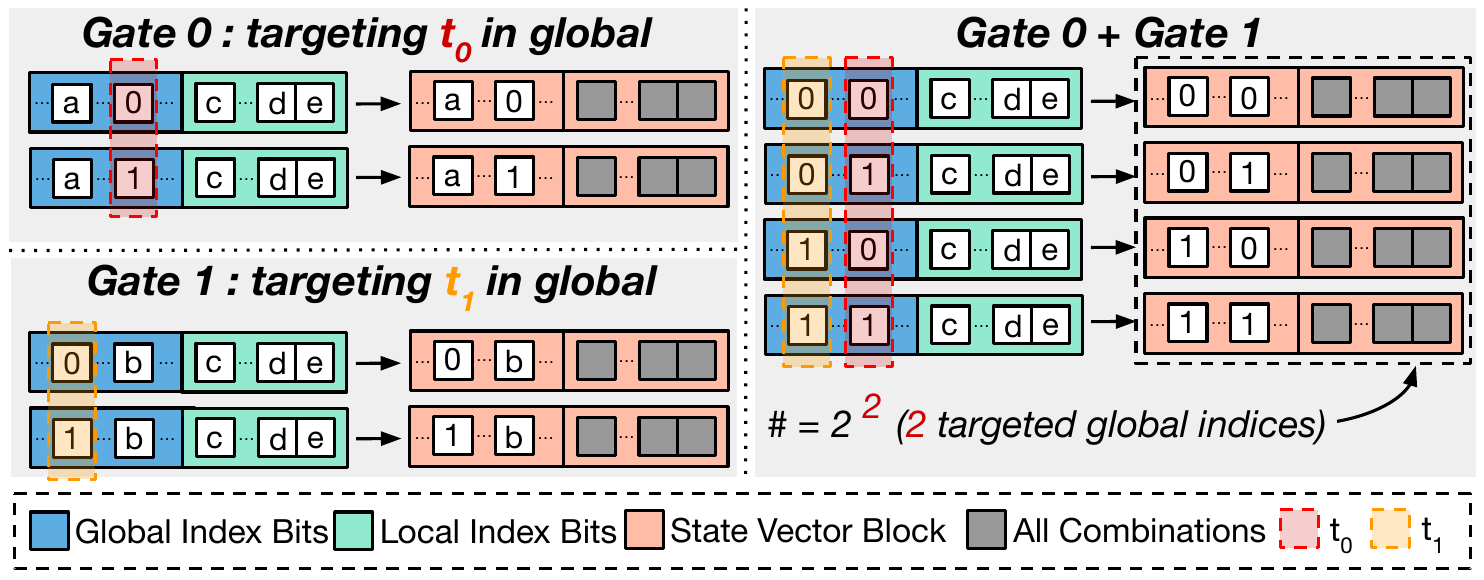}
    \caption{An example of how SV blocks are involved based on target global index changes. The same alphabet denotes the same value (0 or 1).}
    \label{fig:observation}
\end{figure}

\textbf{Insight from the Analysis.} To solve this issue, we carefully analyze Observation in \SEC\ref{sec:problem} and obtain two important findings.
(1) For multiple gates targeting the local index set, we can apply them all after decompressing the corresponding SV block because every amplitude in this block can find its corresponding pair within the same block. (2) For multiple gates targeting the global index set, since different gates may require pairs of different SV blocks, we can involve a few more SV blocks to make the multi-gate application possible and balance the far-reach of pairs. An example of this is shown in Figure \ref{fig:observation}: when two gates targeting different global indices are applied, we can include more SV blocks to ensure that the pairs of amplitudes needing updating can still be found within these SV blocks. The number of SV blocks involved is two to the power of the number of targeted global indices. Insight is drawn from above findings:
\begin{boxI}
\textbf{Insight}: If all the gates in the circuit target the local index or a few specific global indices, then the state vector update can be done for all the gates with one decompression.
\end{boxI}

\begin{figure}[t]
    \centering
    \includegraphics[width=\linewidth]{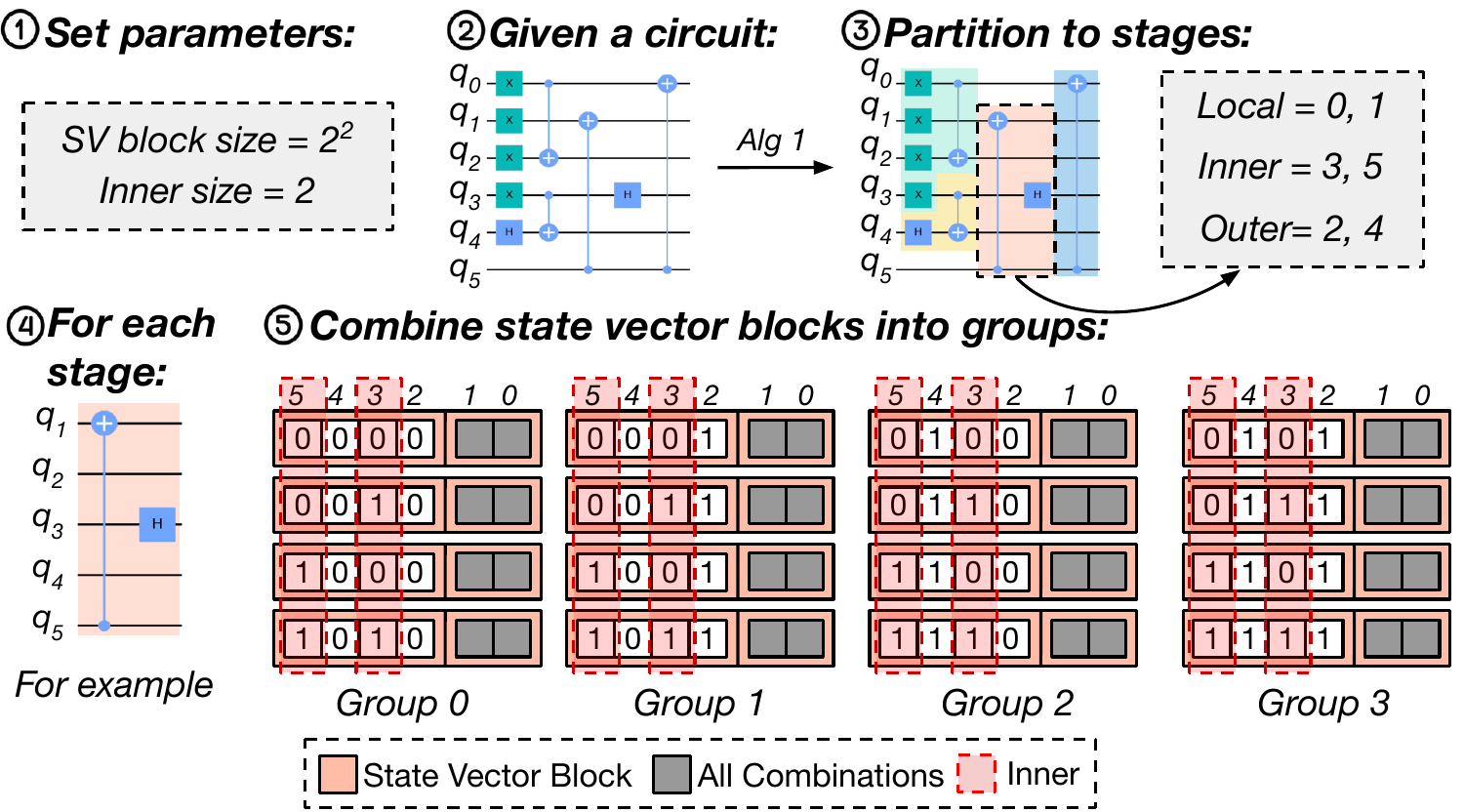}
    \caption{An example of the proposed circuit partition process.}
    \label{fig:circuit_partition_2}
\end{figure}

\textbf{How to make the circuit consist only of gates targeting certain indices?} We find that if we partition the circuit into multiple \uline{stages} where the number of global indices targeted by the involved gates in a stage is limited, then within such stage, all the gate operations can be performed using the same SV block access pattern. Therefore, we propose a circuit partition algorithm to partition the circuit into stages given a pre-defined limit of number of global indices. Details of this approach can be found in Algorithm \ref{alg:circuit_partition}. Specifically, we define global indices that appear within a stage as \uline{inner indices} and other global indices that do not as \uline{outer indices}. After the user specifies the SV block size and the inner size, the algorithm runs \textbf{offline} for a given circuit. For each stage, we add one gate at a time from the input circuit (Line 11) until the number of global indices in the stage reaches a threshold (Lines 7-9). We repeat this process until the circuit is fully traversed (Line 4). Note that the minimum number of inner indices must be two (Line 3). This requirement stems from the structure of quantum circuits, consisting of single- and double-qubit gates. Ensuring at least two inner indices is crucial for effective circuit partitioning when a double-qubit gate's target qubits both fall within the global indices.

\begin{algorithm}[b]
\footnotesize
\caption{Proposed circuit partition method.}
\begin{algorithmic}[1] 
    \Require{circuit, SV block size, inner size}
    \Ensure{stages}
    \State stages = [] 
    \State current stage = [] 
    \State threshold = max(inner size, 2) \Comment{\textcolor{blue}{2 for double-qubit gates}}
    \While{i < number of gates in circuit}
    \State current gate = circuit[i]
    \State query the global indices in [current stage + current gate]
    \If{exceed the threshold} \Comment{\textcolor{blue}{Partition current stage}}
        \State add current stage to stages 
        \State current stage = [] \Comment{\textcolor{blue}{Clear current stage}}
    \EndIf
    \State add current gate to current stage
    \State i++
    \EndWhile
    \If{current stage not empty} 
        \State add current stage to stages
    \EndIf
\end{algorithmic}
\label{alg:circuit_partition}
\end{algorithm}

An example of this process is depicted in Figure \ref{fig:circuit_partition_2}. In this example, we partition the circuit into four stages with Algorithm \ref{alg:circuit_partition}. For this 6-qubit (\(n=6\)) circuit, the local index size is 2 (\(b=2\)), and the global index size is 4 (\(c=4\)). In the example stage from the step 4 in Figure \ref{fig:circuit_partition_2}, indices 3 and 5 are the inner indices of this stage, while 2 and 4 are the outer indices. All the gate operations in this stage only involve the SV blocks with the same outer indices. 
We call this set of SV blocks an \uline{SV group}; there are a total of 4 groups in this example. Each group can be updated independently.

With this design, each stage requires only one compression and one decompression operation, significantly reducing the frequency of compression. For instance, in the simulation of a 33-qubit QFT circuit, our approach can decrease the number of compression occurrences from 2,673 (i.e., the number of gates) to just 28 (i.e., the number of stages). This substantially increases the final result's fidelity and also improves overall performance.

\subsection{Transfer-Concealed Workflow}
\label{subsec:pipeline-design}
On one hand, to maximize qubit support, it is beneficial to store compressed state-vector (SV) blocks in the larger CPU memory (e.g., 16GB to 512GB) compared to GPU memory (e.g., 4GB to 80GB). On the other hand, GPU-based simulators outperform CPU-based ones due to high parallelism ideal for matrix multiplication. Therefore, \thiswork{} leverages both CPU and GPU: compressing SV blocks in CPU memory and assigning state vector updates to GPUs. This design, however, requires frequent CPU-GPU memory transfers, complicating block-wise state vector updates.

\begin{figure}[t]
    \centering
    \includegraphics[width=\linewidth]{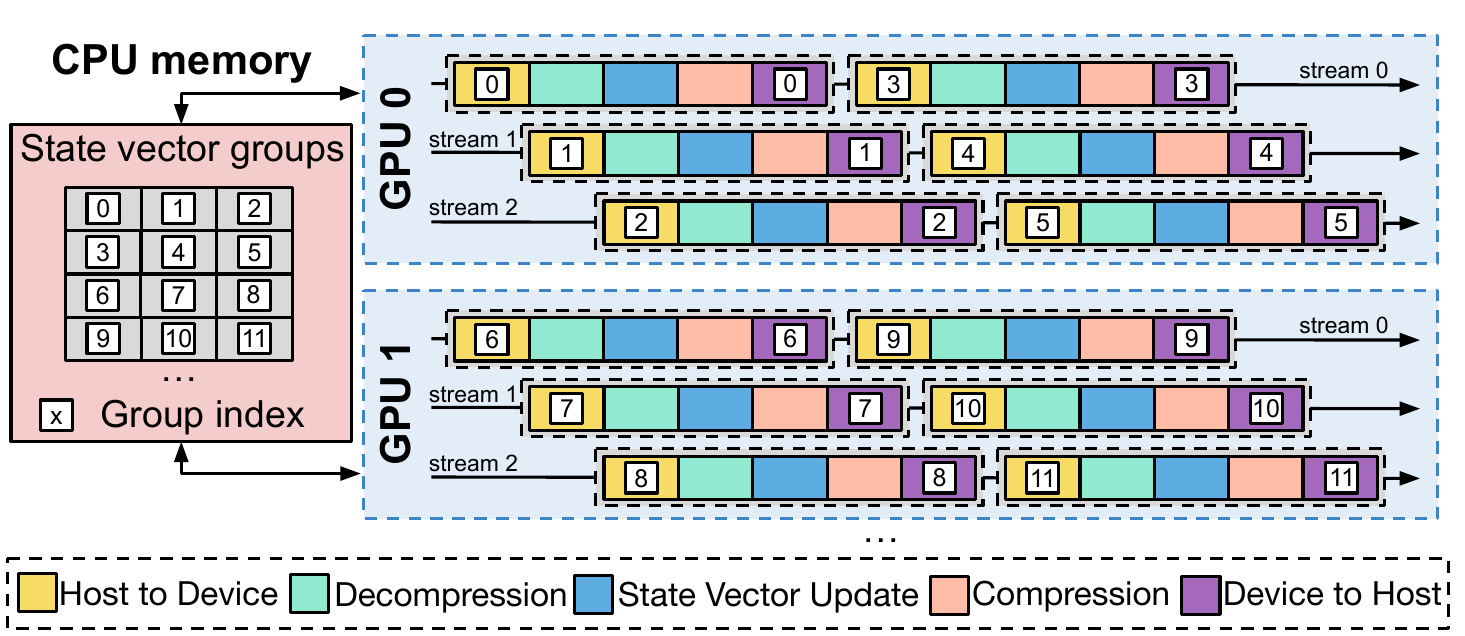}
    \caption{A demonstration of our multi-stream pipeline design.}
    \label{fig:stream}
\end{figure}

\textbf{Pipeline design.} To resolve this issue, we propose a memory transfer and computation overlapping pipeline. As described in \SEC\ref{subsec:hardware-sv-partition}, the simulation is divided into discrete, independent tasks called SV groups, allowing for more modular and efficient processing. This characteristic is utilized to overlap kernel executions with data transfers (as mentioned in \SEC\ref{subsec:gpu}, GPUs can perform memory copy operations and kernel execution concurrently). A demonstration of this pipeline design is shown in Figure \ref{fig:stream}: each SV group undergoes a sequence of operations including host-to-device memory copy, decompression, state vector updating, compression, and device-to-host memory copy. These operations are scheduled on the same CUDA stream to maintain the correct execution order. Additionally, operations for different SV groups are scheduled to each CUDA stream repeatedly, facilitating the overlap of overall processes. Moreover, kernel executions can also be overlapped by the GPU scheduler to fully leverage the computing resources in the GPUs. This strategy efficiently overlaps memory operations and kernel execution, enhancing overall performance.


\textbf{Multi-GPU parallelization.} Since the simulation process is divided into independent tasks by our circuit partition, different GPUs can simultaneously process distinct SV groups of SV blocks. This enables native support for concurrency at the inter-GPU level in \thiswork. As shown in Figure \ref{fig:stream},  each GPU handles partial SV groups and processes them locally without GPU-to-GPU communication. Note that the throughput of multi-GPU parallelization is bounded by the PCIe bandwidth, as all data transfer between the CPU and GPUs occurs through PCIe. When memory movement is intensive, it can cause a starvation problem for GPUs (evaluated in Section~\ref{subsec:other_eval}).

Note that in the beginning of the simulation, the state vector is initialized to a standard base state (the first element is 1, all the others are 0) as a common practice~\cite{li2021qasmbench}. 
When the initial state differs from this standard as the simulation proceeds, a few quantum gates can be used to establish the desired initial state. After partitioning the state vector, all SV blocks, except the first one, consist only of zeros. Therefore, there is no need to compress the same SV block multiple times. During the initial compression, we only need to compress the block with the first element set to one and another block containing all zeros. Then, we can copy the compressed SV block with all zeros multiple times. This approach reduces the (de)compression overhead by one instance.

\subsection{Point-wise Error Control for GPU Compression}
\label{subsec:gpu-compression}
We introduce our proposed GPU point-wise compression error control to ensure that the compression-error propagation in simulation can be bounded in the final results.

It has been proven that GPU lossy compression has much better performance and similar compression ratios compared to CPU compression. To this end, we employ GPU lossy compression in \thiswork{} to minimize the compression overhead. Previous work has demonstrated a lower bound on the fidelity of the state vector when applying a point-wise relative error bound \cite{wu2019full}. Unfortunately, to our knowledge, current SOTA GPU lossy compressors do not support the point-wise relative error bound mode. To address this, we propose the GPU Point-wise Error Compression algorithm. Drawing on previous work by Liang \textit{et al.} \cite{liang2018efficient}, we use a logarithmic transformation to convert point-wise relative error bounds to absolute error bounds.

Specifically, let \( f(x) = \log_2(x) \) be a bijective transformation of the original data point \( x \). Applying an absolute error bound \( b_a \) to \( f(x) \) results in the original data being bounded by a point-wise relative error \( b_r \), as shown by the equation:
\begin{equation}
\frac{|f^{-1}(f(x) + b_a) - x|}{|x|} \leq b_r
\end{equation}

The relationship between \( b_a \) and \( b_r \) can be expressed as:
\begin{equation}
\label{eq:log}
b_a = g(b_r) = \log_{2}(1 + b_r)
\end{equation}
As a result, we can achieve point-wise relative error bounds.

\textbf{Challenges.} Note that the \( \log_2 \) transformation in Equation (\ref{eq:log}) requires positive input values, but satisfying this requirement is a non-trivial task. A common method is to convert negative values to absolute values before applying the \( \log_2 \) transformation and use an array to record their indices. However, this approach would significantly lower the overall compression ratio due to the extra space for the index array, potentially even leading to data size inflation.

\textbf{Our solution.} To address this challenge, we propose an algorithm that avoids using an index array to record the negative values. We detail this algorithm in Algorithm \ref{alg:compression} (the decompression process is simply the inverse). Specifically, we use a bitmap to store the sign of each number in the original array (Line 1), designating 0 for positive values and zeros (Line 8), and 1 for negative values (Line 5). Then, we convert the negative values to their additive inverse (Line 6) and apply the log transformation (Line 10). Subsequently, we apply lossy compression with absolute error-bounded mode to the data to achieve point-wise error control (Line 15).

Note that based on our observations, bitmaps frequently exhibit long sequences of repeated 0-bits or 1-bits, indicating that the sign of the state vector is often repeated over extensive distances. To address this redundancy, we propose a pre-scan of the bitmap (Line 16). Specifically, the bitmap is partitioned into chunks, within which CUDA's warp-level fast scan functions, \_\_ballot\_any and \_\_ballot\_all, are employed. These functions, optimized by register direct data exchange, rapidly assess large bitmap chunks to determine if all bits within a chunk are all-0 or all-1. The results are recorded, and redundant all-0 or all-1 chunks are removed. The remaining data is finally compressed using an additional lossless encoding method (Line 17). This approach not only increases the compression ratio but also enhances overall compression performance.

\begin{algorithm}[t]
\footnotesize
\caption{GPU point-wise relative error control compression.}
\begin{algorithmic}[1]
    \Require{SV blocks}
    \Ensure{compressed SV blocks, compressed bitmap}
    \State bitmap = []
    \While{i < number of SV blocks} \Comment{\textcolor{blue}{Pipelined in Section 4.2}}
        \While{j < number of elements in SV block}
            \If{SV block[i][j] < 0}
                \State add 1-bit to bitmap \Comment{\textcolor{blue}{1 denotes a negative number}}
                \State SV block[i][j] = -SV block[i][j] \Comment{\textcolor{blue}{Convert to positive}}
            \Else
                \State add 0-bit to bitmap \Comment{\textcolor{blue}{0 denotes a non-negative number}}
            \EndIf
            \State SV block[i][j] = $log_2$(SV block[i][j]) \Comment{\textcolor{blue}{Convert to log scale}}
            \State j++
        \EndWhile
        \State i++
    \EndWhile
    \State lossy encode (SV block) 
    \State pre-scan(bitmap)
    \State lossless encode (bitmap)
\end{algorithmic}
\label{alg:compression}
\end{algorithm}

\subsection{Two-Level Memory Management}
\label{subsec:fall-back}
The point-wise error-bounded lossy compression introduced in \thiswork{} raises a potential issue: no sufficient memory guarantee for simulation due to variable compression ratios during the simulation. To address this, we propose a two-level memory management system. Specifically, if the main memory is insufficient, the machine's storage component is employed as a fallback strategy to support the simulation.

\textbf{Challenges.} A couple of reasons make this solution challenging:
1. Data transfer from the storage to the GPU requires an intermediate step of involving CPU memory, needing additional memory space as a temporary buffer for SV blocks from the storage.
2. Moving SV blocks from the storage to CPU memory and then to GPU memory generates significant latency, degrading the overall simulation performance.

\textbf{Our solution.} To address these challenges, we employ the GDS technology (as introduced in \SEC\ref{subsec:gpu}) to enable direct memory access between GPU and storage, leveraging the Direct Memory Access (DMA) engine. This method bypasses the potential CPU bounce buffer that traditionally is used as an intermediary for transferring memory between the storage and GPU global memory. Utilizing GDS not only conserves CPU memory—heavily employed for storing compressed SV blocks—but also minimizes CPU overhead. This application of GDS in our design thus enhances \thiswork{}'s capacity to handle larger quantum simulations more robustly.

During the simulation, if \thiswork{} detects that there is insufficient memory for an upcoming compressed SV block, it calls the cuFile APIs \cite{cufile} to directly save this chunk to the storage via GDS. Our evaluation (\S~\ref{sec:eval}) indicates that the performance drop with two-level management is not significant (i.e., 0.7\% on average), highlighting our efficient design.

\section{Experimental Evaluation}
\label{sec:eval}

\subsection{Experimental Setup}
\label{sub:evalsetup}

\textbf{Machines.}  
Due to the administrative privileges required for driver support for the GDS technique, we conduct our evaluation primarily using the following two machines:  

\underline{Machine 1}: A workstation equipped with a 28-core Intel Xeon Gold 6238R CPU at 2.20GHz and two NVIDIA GTX A4000 GPUs (40 SMs, 16 GB each), along with 128 GB DDR4 memory. This workstation runs Ubuntu 20.04.5 and CUDA 12.3.107. It also includes a Samsung 870 EVO MZ-77E4T0E SSD with a capacity of 4 TB and a SATA 6Gb/s interface. The GPUs in this workstation are connected via PCI Express 4.0.

\underline{Machine 2}: To evaluate multi-GPU performance speedup, we also include a node from an HPC cluster, which includes a 64-core AMD EPYC 7713 CPU at 2.00GHz and four NVIDIA Ampere A100 GPUs (108 SMs, 40GB each). This system has 256 GB DDR4 memory and runs CentOS 7.4 with CUDA 12.2.91. The GPUs are interconnected using NVLink.

\textbf{Software.}  
We implement \thiswork{} based on SV-Sim \cite{li2021sv}, primarily because SV-Sim (already merged into NWQSim \cite{suhsimulating}) is an open-source platform with active maintenance. Furthermore, we base our compression on bitcomp from NVCOMP \cite{nvcomp}, as bitcomp excels among GPU lossy compressors for its exceptional compression throughput and ratio. Bitcomp integrates both lossless mode and lossy mode. We use lossless mode for bitmap and lossy mode for data. We use a point-wise relative error bound of \(10^{-3}\), as this provides a balanced compression ratio and fidelity.

\textbf{Baselines.}  
We compare \thiswork{} with the following baselines: SV-Sim \cite{li2021sv}, Qiskit-Aer \cite{qiskit}, cuQuantum Appliance \cite{cuquantum}, and HyQuas \cite{zhang2021hyquas}. Each of these supports GPU-based state-vector simulation. Additionally, we include a comparison with another state-vector simulation work utilizing compression, referred to as SC19-Sim \cite{wu2019full}. However, as the implementation of SC19-Sim is not publicly available, we developed a prototype of SC19-Sim with SV-Sim and SZ2 \cite{sz2, liang2018error}. A detailed comparison is presented in Table \ref{tab:comparison}.

\begin{table}[ht]
    \centering
    \caption{Comparison of Different State Vector Simulators}
    \resizebox{\columnwidth}{!}{
    \begin{tabular}{llcc}
    \toprule
    \textbf{Existing State} & \textbf{State Vector} & \textbf{GPU} & \textbf{Use} \\
    \textbf{Vector Simulators} & \textbf{Location} & \textbf{Updating?} & \textbf{Compression?} \\ \midrule 
    Qiskit & CPU+GPU & \cmark & \xmark \\ 
    SV-Sim & GPU & \cmark & \xmark \\ 
    HyQuas & GPU & \cmark & \xmark \\ 
    cuQuantum & GPU & \cmark & \xmark \\ 
    SC19-Sim & CPU & \xmark & \cmark \\ 
    \textbf{\thiswork{}} & CPU & \cmark & \cmark \\
    \bottomrule
    \end{tabular}}
    \label{tab:comparison}
\end{table}

\textbf{Benchmark Circuits.}  
We select eight quantum algorithms from NWQBench \cite{li2021qasmbench}. This suite includes quantum circuits with qubit numbers ranging from 23 to 33 and gate numbers from 24 to 3010. The selected circuits are cat\_state, cc, ising, qft, bv, qsvm, ghz\_state, and qaoa.

\subsection{Evaluation of Supported Qubit Number}
\label{sub:eval-performance}

We begin by assessing the maximum supported number of qubits across different simulators on Machine 1, as shown in Table \ref{tab:max_qubit_numbers}.
Our evaluations indicate that \thiswork{} can support up to 42 qubits, significantly exceeding other counterparts, which support an average of 30 qubits.
This capacity goes beyond some entire HPC clusters under normal simulation conditions. Note that with the help of an SSD (assuming SSDs as an extral external storage space), \thiswork{} can reach up to 47 qubits, which is close to the capacity of the Frontier HPC cluster at 48 qubits \cite{atchley2023frontier}, and 14 more than other simulators. Note that the supported number of qubits varies due to the unpredictable compression ratio.

\begin{table}[t]
    \centering
    \caption{Maximum Qubit Numbers for Different Simulators on Machine 1.}
    \resizebox{\columnwidth}{!}{
    \begin{tabular}{lccccc}
        \toprule
        \textbf{Algorithm} &\makecell{\textbf{Qiskit} } &\makecell{\textbf{cuQuantum} }& \makecell{\textbf{SV-Sim}} & \makecell{\textbf{HyQuas}} & \makecell{\textbf{\thiswork}} \\
        \midrule
        cat\_state & 33 & 31 & 26 & 29 & \textbf{42} \\
        cc & 30 & N/A & 26 & 29 &\textbf{37} \\
        ising & 33 & 31 & 26 & 29 &\textbf{35}  \\
        qft & 33 & 31 & 26 & 29 &\textbf{36}  \\
        bv & 33 & 31 & 26 & 29 &\textbf{42}  \\
        qsvm & 33 & 31 & 26 & 29 &\textbf{35}  \\
        ghz & 33 & 31 & 26 & 29 &\textbf{42}  \\
        qaoa & 29 & 31 & 26 & 29 &\textbf{35} \\
        \bottomrule
    \end{tabular}}
    \label{tab:max_qubit_numbers}
\end{table}

\begin{figure}[b]
    \centering
    \includegraphics[width=\linewidth]{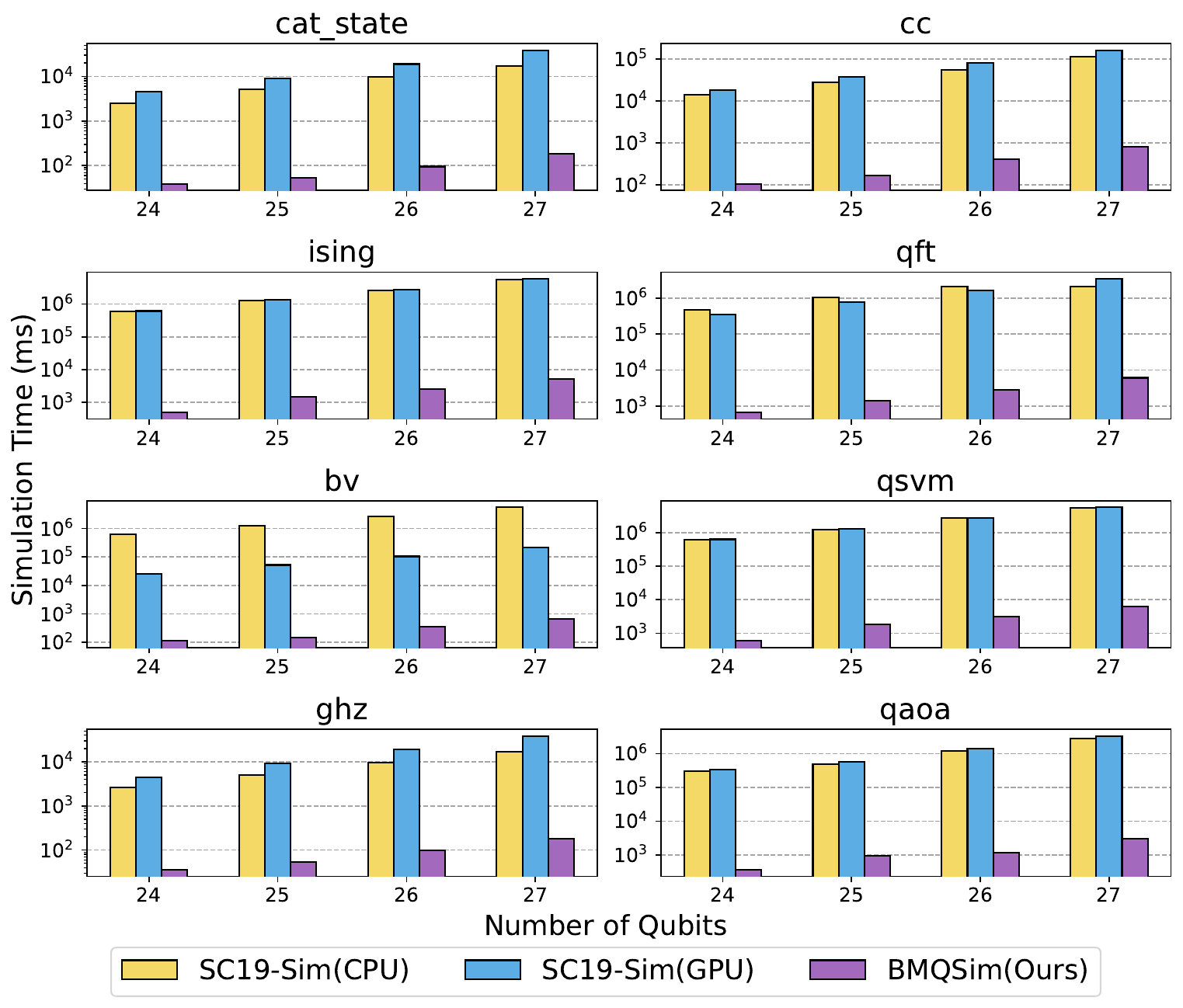}
    \caption{Simulation time of SC19-Sim (CPU/GPU) and \thiswork{}.}
    \label{fig:performance_memory_comparison}
\end{figure}

\subsection{Comparison with SC19-Sim}

We then compare \thiswork{} with another compression-based state-vector simulation, SC19-Sim \cite{wu2019full}, to demonstrate the high-performance and high-fidelity advantages of our work.

Since SC19-Sim is not open-source, we implemented a prototype based on SV-Sim \cite{li2021sv} with the fastest compression technique, solution B, in their paper \cite{wu2019full}. For a fair comparison, we implemented both a CPU version as described in the SC19-Sim \cite{wu2019full} paper and a GPU version using the same compression technique but utilizing GPUs to update the state vectors. We ran this evaluation on Machine 1.

\textbf{Simulation Time.} We begin by comparing the simulation time. The results are shown in Figure \ref{fig:performance_memory_comparison}. Our findings indicate that \thiswork{} outperforms both versions of SC19-Sim under all configurations. The average speedup of \thiswork{} compared to SC19-Sim (CPU) and SC19-Sim (GPU) is $1385\times$ and $539\times$, respectively. This significant performance boost is attributed to the low compression frequency, finely pipelined workflow, and high-performance GPU compression. 
Note that in some cases, the SC19-Sim CPU version outperforms the SC19-Sim GPU version. This anomaly is due to the basic solution implemented in SC19-Sim that does not overlaps the data transfer and kernel execution. This results a huge overhead in the memory movement between the CPU and GPU. In contrast, our work leverages a pipeline design to minimize the overhead of data transfer and gain significant performance improvement (evaluated in Section \ref{subsec:overhead}).

\begin{figure}[t]
    \centering
    \includegraphics[width=\linewidth]{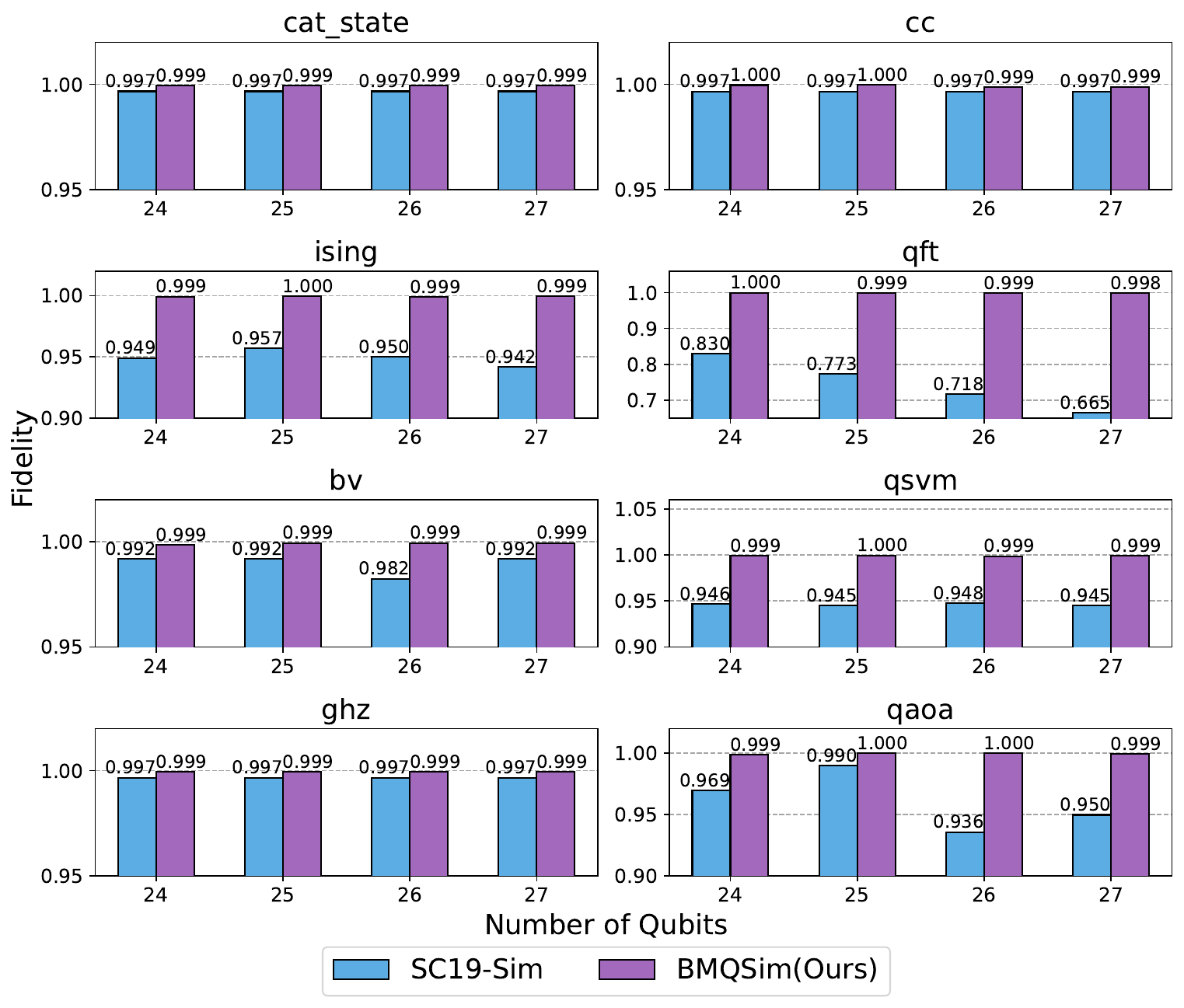}
    \caption{Fidelity of SC19-Sim and \thiswork{} (higher values better fidelity).}
    \vspace{-4mm}
    \label{fig:fidelity}
\end{figure}

\textbf{Fidelity.} Next, we evaluate the fidelity of simulation results. Fidelity is the most important metric for determining the authenticity of final quatum state. It indicates the similarity between the ideal output state and the simulated state, with values ranging from 0 to 1, where higher is better. The fidelity of our simulations is calculated using the equation: \( Fidelity = |\langle\psi_{ideal}|\psi_{sim}\rangle| \), where \( \psi_{ideal} \) is the ideal output state from SV-Sim and \( \psi_{sim} \) is the state produced by the tested lossy-compression enabled simulation. Our results show that \thiswork{} achieves a fidelity greater than 0.99 across all configurations, which is higher than SC19-Sim, particularly for deep circuits. For instance, \thiswork{} achieves $1.35\times$ higher fidelity on average compared to SC19-Sim in the qft circuit.

\subsection{Evaluation of Memory Consumption}

\begin{figure*}[t]
    \centering
    \includegraphics[width=.95\linewidth]{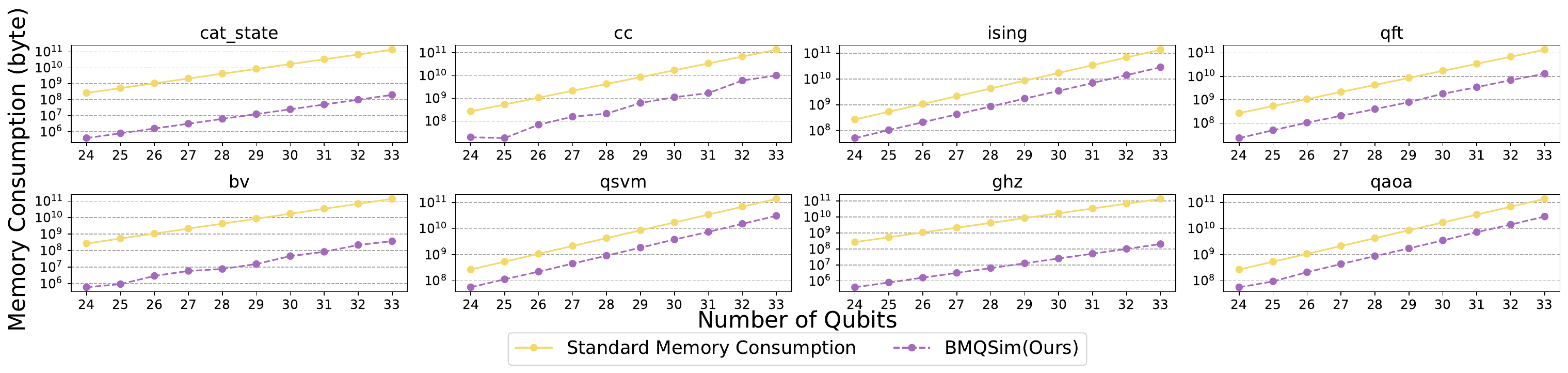}
    \caption{Memory consumption of \thiswork{} compared to the memory required for normal state vector simulation, denoted as standard memory consumption.}
    \label{fig:memory_consumption}
\end{figure*}

We present a memory consumption comparison between \thiswork{} and the standard for state vector simulation, which is \(2^{n+4}\) bytes, where \(n\) denotes the number of qubits, as shown in Figure \ref{fig:memory_consumption}. The (de)compression is performed once for each circuit stage. We consider the maximum memory consumption across all stages in the circuit as the final memory consumption of the simulation. Extremely low memory usage is observed for cat\_state, bv, and ghz\_state, with average memory reductions of 678.61 times for cat\_state, 424.77 times for bv, and 678.52 times for ghz\_state. Other circuits also maintain significant memory reductions, averaging 15.50 times for cc and 10.54 times for qft.

Note that in most cases, system memory is sufficient for simulation. Thus, to evaluate the two-level memory management design that uses SSD storage as a backup plan for simulation, we limit the memory space of Machine 1 to 8 GB and run the same evaluation. We find that the SSD is leveraged only when the qubit number is larger than 32 qubits for some circuits. For example, the ising circuit stores 39\% and 70\% of its SV blocks in the SSD with qubit numbers 32 and 33, respectively.

\subsection{Evaluation of Simulation Time}
\begin{figure*}[t]
    \centering\includegraphics[width=.97\linewidth]{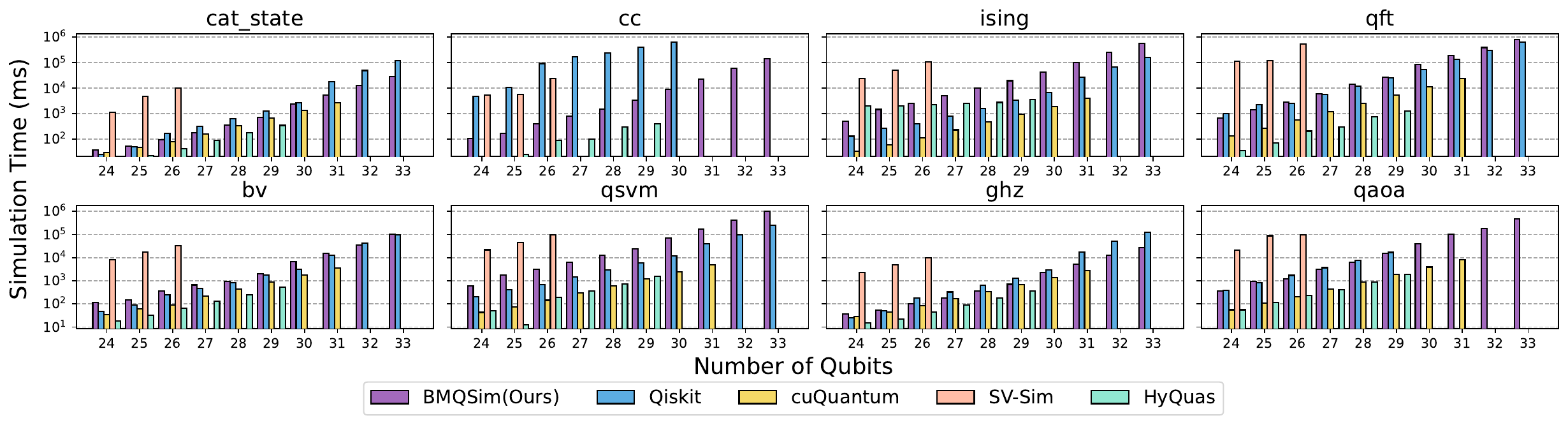}
    \caption{Simulation time of different solutions on various quantum circuits and qubit numbers (missing bars indicate memory allocation errors).}
    \label{fig:simulation_times}
\end{figure*}

Next, we evaluate the simulation time of \thiswork{} compared with other baselines on Machine 1, as shown in Figure \ref{fig:simulation_times}.

Compared to SV-Sim, \thiswork{} offers significant performance improvements. When NVLink is not available, SV-Sim experiences substantial overhead from GPU-to-GPU communication, resulting in the longest simulation times across all settings. In contrast, \thiswork{} partitions the circuit into stages, dividing the simulation into independent local jobs on GPUs, which eliminates the GPU-to-GPU communication, resulting in an average performance speedup of $75\times$.

In most cases, \thiswork{} achieves similar simulation times to the Qiskit-Aer GPU simulator. For instance, the simulation time ratio of \thiswork{} to Qiskit-Aer is 0.99 and 1.05 for qsvm and qft on average, respectively. This demonstrates that \thiswork{} has optimized the simulation process to perform on par with the SOTA GPU simulator from industry. It is important to note that Qiskit-Aer utilizes both CPU and GPU memory for storing the state vector and prioritizes GPU memory based on our evaluation. Consequently, there is a significant drop in performance when the qubit number increases from 30 to 31, as the GPU memory becomes insufficient, causing a fallback to combined memory.

Despite the improvements, both cuQuantum and HyQuas still outperform \thiswork{} in most cases. This performance disparity is primarily due to the SV-Sim backend on which \thiswork{} is based. HyQuas, with its series of performance optimizations, achieves the best performance among all simulators, being $12\times$ faster than \thiswork{} on average. However, this performance comes at the cost of higher memory consumption, which limits HyQuas's supported qubit number compared to other GPU simulators. CuQuantum, tested using the backend integrated in qsim \cite{quantum_ai_team_and_collaborators_2020_4023103}, achieves approximately $9\times$ speedup compared to \thiswork{}. However, cuQuantum is not an open-source tool and only supports the float32 data type. This inherent characteristic renders it faster than all the other evaluated simulators using float64 data points.

Compared to these well-optimized works, the advantage of \thiswork{} lies in its ability to support a considerable larger number of qubits. Given the popularity and acceptance in the community, \thiswork{} offers comparable simulation time with industry-level simulators like Qiskit with significantly more supported qubits. 

\begin{figure*}[h!]
    \centering\includegraphics[width=.97\linewidth]{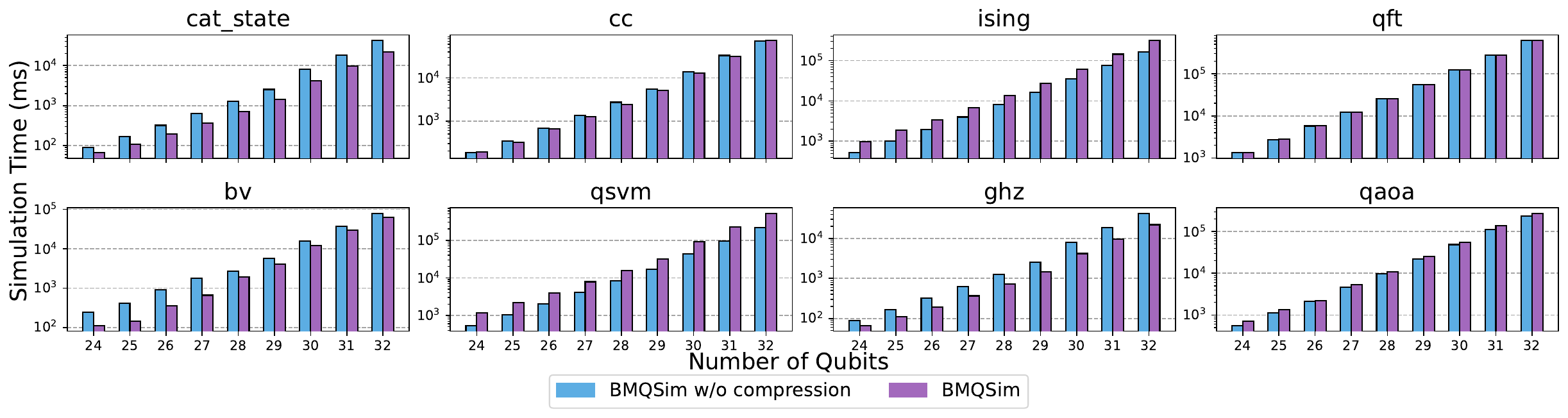}
    \caption{Compression overhead of \thiswork{} with different numbers of qubits.}
    \label{fig:breakdown_analysis}
\end{figure*}

\subsection{Compression Overhead Analysis}
\label{subsec:overhead}
We further evaluate the compression overhead of our design by comparing it with the version of \thiswork{} without compression, as shown in Figure \ref{fig:breakdown_analysis}. For this evaluation, we use a single A4000 GPU in Machine 1 to reduce the impact of other overhead on the evaluation results. The results illustrate that, thanks to our circuit partitioning and pipeline design optimizations, the compression overhead is minimal compared to the version without compression. Notably, in some cases, \thiswork{} even outperforms the no-compression version. This is because, although compression adds overhead to the simulation process, it also reduces memory copy time due to the smaller size of the compressed SV blocks. When the compression ratio is high, as in the cases of the \textit{cat\_state}, \textit{bv}, and \textit{ghz} algorithms, the data copy overhead becomes negligible, enabling \thiswork{} to outperform the version without compression. Overall, the compression technique contributes positively to the simulation time and leads to a $9\%$ speedup on average. In comparison, compression accounts for approximately $61\%$ on average of the SC19-Sim simulation time, demonstrating that our work significantly lifts compression overhead.

\subsection{Pipeline Design Analysis}
\begin{figure}[ht]
    \centering\includegraphics[width=.96\linewidth]{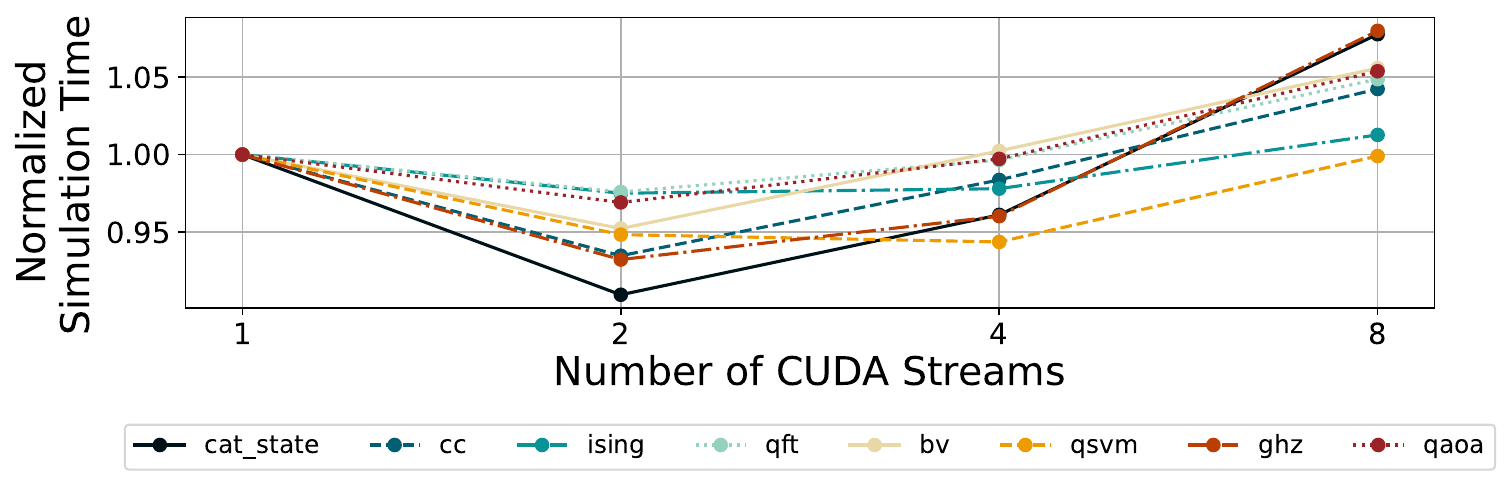}
    \caption{Impact of CUDA stream number in our pipeline design.}
    \label{fig:cuda_streams_latency}
\end{figure}

We also evaluate the impact of different CUDA stream numbers and present the results on Machine 1 in Figure \ref{fig:cuda_streams_latency}. We fix other parameters, such as the SV block size and inner size, to isolate the impact of the stream number. When the CUDA stream number is set to 1, it represents the version of \thiswork{} without pipeline optimization. Our findings indicate that, in most cases, the highest speedup is achieved when the stream number is set to 2. Although the speedup is not as significant with a stream number of 4, some improvement is still observed. However, when the stream number reaches 8, the pipeline version becomes slower than the sequential version. This is due to the stream context switch overhead outweighing the benefits brought by pipeline speedup.

\subsection{Other Evaluations}
\label{subsec:other_eval}
Finally, we evaluate other settings, including the GPU number, inner size, SV block size, and circuit partition overhead.

\begin{figure}[ht]
    \centering\includegraphics[width=\linewidth]{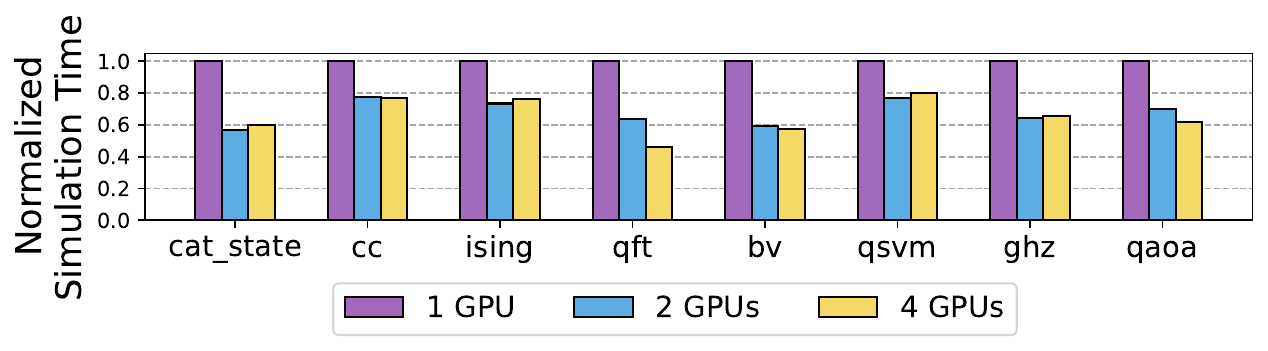}
    \caption{Scalability of \thiswork{} on different algorithms.}
    \label{fig:scale_up}
\end{figure}

\textbf{Multi-GPU Speedup.} To evaluate the scalability of our work, we test it on up to 4 A100 GPUs from Machine 2 with different circuits of 28 qubits, as shown in Figure \ref{fig:scale_up}. In the qft, our work achieves a speedup of $1.7\times$ and $2.3\times$ for 2 GPUs and 4 GPUs, respectively, thanks to the independent SV groups design in \thiswork{}. While the speedup is not significant when the number of GPUs rise from 2 to 4 in some cases due to the CPU and GPU memory transfer rate bounded by the PCIe (as mentioned in Section \ref{subsec:pipeline-design}) and the high overhead of GPU operation launches.

\begin{figure}[ht]
    \centering\includegraphics[width=.9\linewidth]{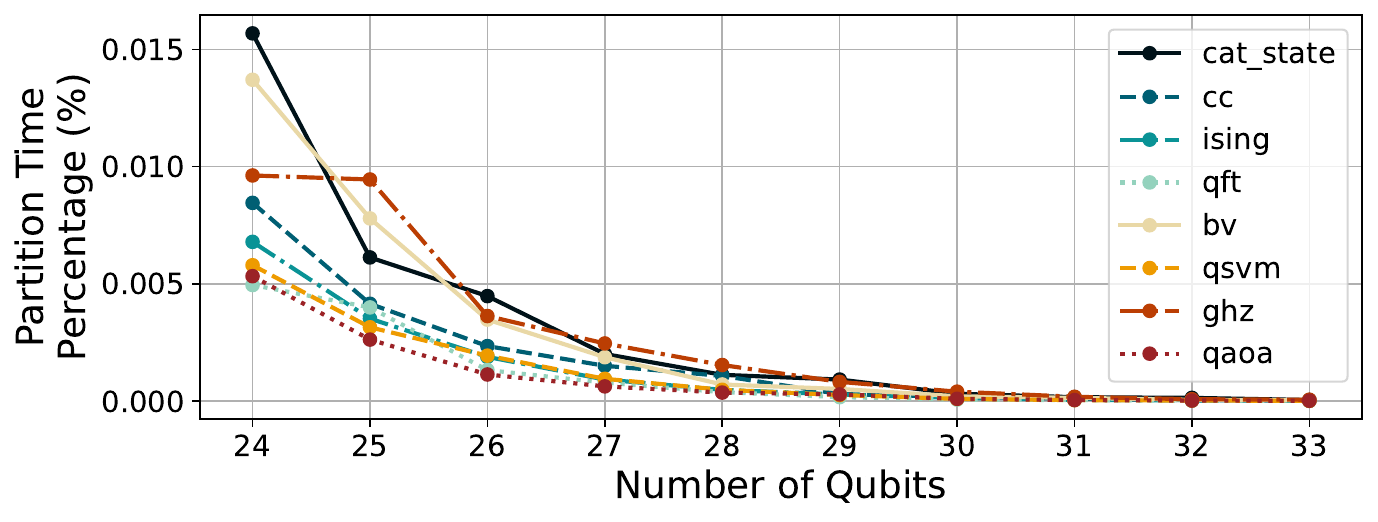}
    \caption{Circuit partition time as a percentage of overall simulation time.}
    \label{fig:partition_time}
\end{figure}

\textbf{Evaluation of Circuit Partition Overhead.} To demonstrate the overhead of the extra circuit partition strategy, we evaluate the percentage of the circuit partition time compared to the end-to-end latency of the simulation process, as shown in Figure \ref{fig:partition_time}. The results indicate that the partition time is negligible compared to the overall simulation time.

\begin{figure}[ht]
    \centering\includegraphics[width=\linewidth]{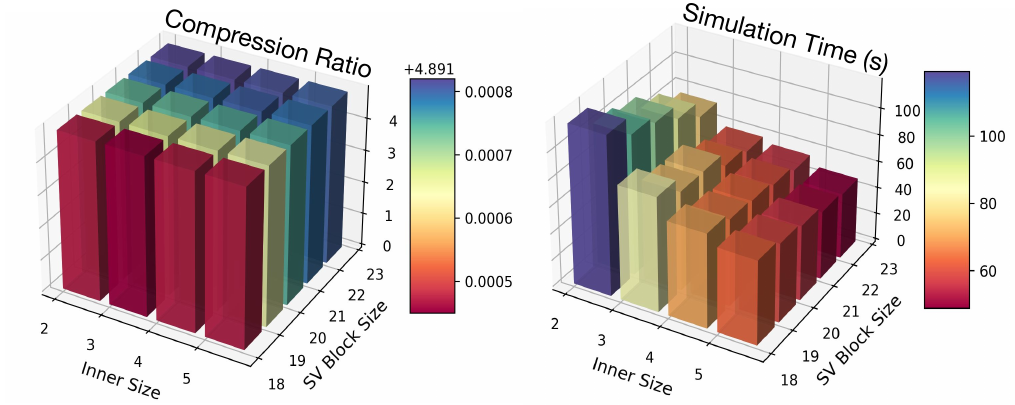}
    \caption{The impact of two system parameters (i.e., inner size and SV block size) on compression ratio (left) and simulation time (right).}
    \vspace{-5mm}
    \label{fig:parameters}
\end{figure}

\textbf{Parameter Tuning.} To evaluate the influences of the inner size and SV block size, we assess the simulation time and compression ratio (the ratio of standard memory to practical memory) with different settings for the 30-qubit qaoa algorithm, as shown in Figure \ref{fig:parameters}. Our findings indicate that the compression ratio does not vary significantly with different inner sizes and SV block sizes. However, the simulation time is shorter with higher inner sizes and SV block sizes. This is because a larger inner size and SV block size result in fewer stages and, consequently, fewer kernel launches.

\section{Conclusion and Future Work}
\label{sec:conclusion}

In this paper, we introduced \thiswork{} to address memory limitations in quantum simulation. By innovatively employing lossy compression and effectively tackling challenges such as low simulation fidelity and high compression overhead, \thiswork{} has successfully enabled the simulation of up to 14 (on average 10) additional qubits under memory constraints, with fidelity over 0.999 in almost all cases.

In the future, we plan to integrating \thiswork with other state-vector simulators such as cuQuantum and Qiskit to improve the performance and increase the usability.

\clearpage
\bibliographystyle{plain}
\bibliography{references}

\end{document}